\documentclass[twocolumn,aps,superscriptaddress,showpacs,floatfix]{revtex4}
\usepackage{amssymb}
\usepackage{amsmath}
\usepackage{graphicx}
\usepackage[normalem]{ulem}
\usepackage[dvips]{color}
\usepackage{bm}
\usepackage{longtable}
\usepackage{array}
\usepackage{etoolbox}
\usepackage{subfigure}
\usepackage{setspace}
\usepackage[colorlinks,linkcolor=blue,anchorcolor=blue,citecolor=blue,urlcolor=blue]{hyperref}

\renewcommand\sout{\bgroup \color{red} \ULdepth=-.5ex \ULset}

\begin{document}
\title{Systematic study of $\alpha$ decay for odd-$A$ nuclei within a two-potential approach}

\author{Xiao-Dong Sun}
\affiliation{School of Math and Physics, University of South China, 421001 Hengyang, People's Republic of China}
\author{Chao Duan}
\affiliation{School of Nuclear Science and Technology, University of South China, 421001 Hengyang, People's Republic of China}
\author{Jun-Gang Deng}
\affiliation{School of Nuclear Science and Technology, University of South China, 421001 Hengyang, People's Republic of China}
\author{Ping Guo}
\affiliation{School of Math and Physics, University of South China, 421001 Hengyang, People's Republic of China}
\author{Xiao-Hua Li\footnote{%
Corresponding author: lixiaohuaphysics@126.com }}
\affiliation{School of Nuclear Science and Technology, University of South China, 421001 Hengyang, People's Republic of China}
\affiliation{Cooperative Innovation Center for Nuclear Fuel Cycle Technology $\&$ Equipment, University of South China, 421001 Hengyang, People's Republic of China}
\affiliation{Key Laboratory of Low Dimensional Quantum Structures and Quantum Control, 410081 Changsha, People's Republic of China}

\begin{abstract}
$\alpha$ decay is usually associated with both ground and low-lying isomeric states of heavy and superheavy nuclei, and the unpaired nucleon plays a key role on $\alpha$ decay. In this work, we systematically studied the $\alpha$ decay half-lives of odd-$A$ nuclei, including both favored and unfavored $\alpha$ decay within the two-potential approach based on the isospin dependent nuclear potential. The $\alpha$ preformation probabilities are estimated by using an analytic formula taking into account the shell structure and proton-neutron correlation, and the parameters are obtained through the $\alpha$ decay half-lives data. The results indicate that in general the $\alpha$ preformation probabilities of even-$Z$, odd-$N$ nuclei are slightly smaller than the odd-$Z$, even-$N$ ones. We found that the odd-even staggering effect may play a more important role on spontaneous fission than $\alpha$ decay. The calculated half-lives can well reproduce the experimental data. 
\end{abstract}

\maketitle

\section{Introduction}
The odd-even staggering effect in nuclear physics is ubiquitous, for example binding energy, abundance of isotopes, excited energy spectrum, moment of inertia, $\alpha$ preformation probability and so on~\cite{Hir05,Del16,Xu99}. It is primarily caused by pairing correlations that is similar to the electronic superconductivity and the unpaired nucleon blocking effect, while comparable contribution comes from the deformed mean field~\cite{Sat98}. Generally, it is believed that nuclei with the even number of protons and neutrons are more stable than the neighboring odd-$A$ and doubly-odd nuclei. However, for some heavy and superheavy nuclei dominated by $\alpha$ decay and fission, this understanding is biased and the effect of unpaired nucleon on the nuclear stability cannot be neglected~\cite{NNDC}. An explicit example is $^{255}$Rf with a half-life of 2.3 second, while the neighboring even-even nuclei are more instable, e.g., spontaneous fission for $^{254, 256}$Rf isotopes take place within 23 microsecond and 6.4 millisecond at average, respectively. Within the Gamow picture, the process of $\alpha$ decay can be qualitatively described as an $\alpha$ cluster preforms in the surface of the decaying nucleus followed by a quantum tunneling of the potential barrier~\cite{Gam28,Yar16}. Among the above two processes, $\alpha$ particle preformation contains more nuclear structure information~\cite{Qi16,Zha08,Xu07,Xu16}. And the penertration depends on the $\alpha$ decay energy and interaction between the $\alpha$ particle and daughter nucleus without obvious odd-even staggers, while the $\alpha$ preformation probability shows odd-even staggering effect~\cite{Xu05}. 

$\alpha$ decay is a common and important decay mode for heavy and superheavy nuclei, which provides many unique nuclear structure information especially for low lying isomeric and ground states of heavy and superheavy nuclei~\cite{Oga15,Hof16,Heb16,Ren12,Wan10,Yan15}. And the $\alpha$ decay and $\alpha$ clustering are very old subjects but still attracting much attention in both theortical and experimental research~\cite{Bet36,Rop98}. The nuclear cluster structure appears in the regions of either light or heavy nuclei~\cite{Del16,Toh01,Var92,Ast10}. Theoretically, microscopic mechanism on how $\alpha$ cluster forms in the surface of heavy nucleus is still an open question, meanwhile some conclusions on the $\alpha$ preformation probability have been obtained to some extent~\cite{Qi16,Zha08,Sei15}. Notably, there is a striking similarity between the tendency of the neutron pairing gaps and the $\alpha$ preformation probability~\cite{Qi16}. The calculations with an effective isospin-dependent contact interaction show that the di-neutron cluster structure in middle and heavy nuclei is related to the strength of pairing correlations in dilute nuclear matter~\cite{Isa08}. Also, a quartet correlations may be of importance for very low density~\cite{Rop98,Elh16}. Phenomenologically, by analyzing the data of $\alpha$ decay, it is founded that the $\alpha$ preformation probability takes local minimum for nuclei with nucleons number near to the magic numbers~\cite{Zha08,Xu07}. Moreover, other factors, e.g., the valence proton-neutron correlation, the $\alpha$ particle angular momentum effect and the deformation of nuclei, also play a role on the $\alpha$ preformation probability~\cite{Kaz03,Sun162,Sei15}. However, because the precise size of decaying nucleus is unknown, by adjusting the radius parameters and keeping the $\alpha$ preformation probability constant, good reproduces of experimental $\alpha$ decay width can also be obtained~\cite{Zde13}. This problem is actually related to the properties of dilute nuclear matter and the definition of nuclear radius. For nuclei with the finite number of protons and neutrons, the active valence nucleon, especially the odd nucleon, plays an important role on the bulk and collective properties, such as the $\alpha$ preformation probability. In order to investigate the odd-even staggering effect in $\alpha$ decay, we systematically calculate the $\alpha$ decay half-lives for odd-$A$ nuclei. 

Recently we have preformed systematic calculation on $\alpha$ decay half-lives and $\alpha$ preformation probability for even-even nuclei based on the isospin dependent nuclear potential~\cite{Sun161}. In this paper, our previous work~\cite{Sun161} is extended to calculate the $\alpha$ decay half-lives of odd-$A$ nuclei. The odd-even staggering effect of $\alpha$ decay half-lives is discussed in detail. The unpaired nucleon plays a key role on the $\alpha$ preformation probability, which resulting in the complexity of $\alpha$ decay especially for unfavored $\alpha$ decay. This paper is organized as follows. Sec. II briefly shows the theoretical framework of the two-potential approach and the analytic expression for the $\alpha$ preformation probability. In Sec. III we show the systematic results of $\alpha$ decay half-lives for odd-$A$ nuclei and discuss the odd-even staggering effect. A summary is given in Sec. IV.

\section{Theoretical framework}

In the two-potential approach~\cite{Gur87}, the total potential between the $\alpha$ particle and the daughter nucleus contains the nuclear and Coloumb potential besides the centrifugal potential for unfavored $\alpha$ decay. The nuclear potential $V_\text{N}(r)$ is determined within a two-body model, which is different from the quasi-molecular shape path proposed by the Generalized Liquid Drop Model~\cite{Guo15}. In the two-body model, the $\alpha$ particle is assumed to preformed at the surface of decaying nucleus, and the strong attractive nuclear interaction can be roughly approximated by a square well potential~\cite{Zde13}. A better description can be obtained by the double-folding integral of M3Y nucleon-nucleon force between the $\alpha$ particle and residual nucleus~\cite{Xu05}. In this work we choose a type of cosh parametrized form for the nuclear potential, which is also successful in describing the rotatinally spaced energies, enhanced E2 transition strengths and so on~\cite{Buc92,Buc89}. The nuclear potential $V_\text{N}(r)$ can be expressed as
\begin{eqnarray}\label{1}
V_\text{N}(r)=-V_0\frac{1+cosh(R/a)}{cosh(r/a)+cosh(R/a)},
\end{eqnarray}
where $V_0$ and $a$ are the depth and diffuseness for the nuclear potential, respectively. In previous work, we have obtained a set of isospin dependent parameters, which is $a=0.5958$ fm and $V_0=192.42+31.059\frac{N-Z}{A}$ MeV~\cite{Sun161}, where $N$, $Z$ and $A$ are the neutron, proton and mass numbers of the daughter nucleus, respectively. Besides, $R$ and $r$ denote the nuclear sharp radius of the parent nucleus and the distance between the $\alpha$ particle and daughter nucleus, respectively. $R$ is empirically given by~\cite{Roy00}
\begin{eqnarray}\label{2}
R=1.28A^{1/3}-0.76+0.8A^{-1/3}.
\end{eqnarray}
The Coloumb potential $V_\text{C}(r)$ is obtained under the assumption of a uniformly charged sphere, which can be written as
\begin{eqnarray}\label{3}
V_\text{C}(r)= \left \{
\begin{aligned}
\frac{Z_\text{d}Z_\alpha e^2}{2R}[3-(\frac{r}{R})^2],~~~r<R,\\
\frac{Z_\text{d}Z_\alpha e^2}{r},~~~~~~~~~~~~~~~~r>R,
\end{aligned}
\right.
\end{eqnarray}
where $Z_\text{d}$ and $Z_\alpha$ are proton numbers of the daughter nucleus and $\alpha$ particle, respectively. In addition, unfavored $\alpha$ decay has $l \neq 0$, where $l$ is the orbital angular momentum taken away by the $\alpha$ particle. In this $\alpha$ decay mode, the centrifugal potential $V_\text{l}(r)$ plays a role on the total potential, which is given by
\begin{eqnarray}\label{4}
V_\text{l}(r)=\frac{l(l+1)\hbar^2}{2\mu r^2},
\end{eqnarray}
where $\mu$ denotes the reduced mass of the two-body model between $\alpha$ particle and the daughter nucleus in center of mass system.

The $\alpha$ decay half-life $T_{1/2}$ can be calculated by the WKB approxination. It can be written as
\begin{eqnarray}\label{5}
T_{1/2}=\frac{\hbar ln2}{\Gamma}=\frac{ln2}{\lambda},
\end{eqnarray}
where the $\alpha$ decay constant $\lambda$ or decay width $\Gamma$ depends on the $\alpha$ particle preformation probability $P_{\alpha}$, the penetration probability $P$ and the normalized factor $F$. Within the two-potential approach framework~\cite{Gur87}, the decay width can be expressed as
\begin{eqnarray}\label{6}
\Gamma=\frac{P_{\alpha}FP}{4\mu}.
\end{eqnarray}
The normalized factor $F$ represents the collision probability or assault frequency. It satifies the expression
\begin{eqnarray}\label{7}
F\int_{r_1}^{r_2}\frac{dr}{2k(r)}=1,
\end{eqnarray}
where $r_1$, $r_2$ and the following $r_3$ denote the classical turning points, which satisfy the conditions $V(r_1)=V(r_2)=V(r_3)=Q$. The middle turning point $r_2$ is slightly larger than the nuclear sharp radius $R$ of the parent nucleus. $k(r)=\sqrt{\frac{2\mu}{\hbar^2}\mid Q_\alpha-V(r)\mid}$ is the wave number. $V(r)$ and $Q$ are $\alpha$-core potential and $\alpha$ decay energy, respectively. The penetration probability $P$ can be calculated using
\begin{eqnarray}\label{8}
P=exp(-2\int_{r_2}^{r_3}k(r)dr).
\end{eqnarray}

At last, we focus on the most unclear concept, the $\alpha$ particle preformation in the decaying parent nucleus. Due to the absence of effictive method to calculate the $\alpha$ preformation probability, which may be obtained by
\begin{eqnarray}\label{9}
P_\alpha=P_0\frac{T_{1/2}^\text{calc}}{T_{1/2}^\text{expt}},
\end{eqnarray}
where $T_{1/2}^\text{calc}$ is calculated by eq. (5) with $P_\alpha=1$. According to the density-dependent cluster model~\cite{Xu05}, $P_0$ is the average $\alpha$ preformation factor for a certain kind of $\alpha$ decay, such as $P_0=0.7$ for even-even nuclei, $P_0=0.35$ for odd-$A$ nuclei, and $P_0=0.18$ for doubly-odd nuclei. And it takes as $P_0=0.35$ in this work. The variation of $\alpha$ preformation probability can be estimated by the analytic formula~\cite{Zha08,Guo15,Sun161}, which can be described by the following expression
\begin{eqnarray}\label{10}
log_{10}P_\alpha=&a+b(Z-Z_1)(Z_2-Z)+c(N-N_1)(N_2-N)\nonumber\\
&+dA+e(Z-Z_1)(N-N_1),
\end{eqnarray}
where $Z$, $N$ and $A$ are the proton, neutron and mass numbers of parent nucleus. $Z_1$ and $Z_2$ ($N_1$ and $N_2$) are the proton (neutron) magic numbers around $Z$ ($N$). $a$, $b$, $c$, $d$ and $e$ are the adjustable parameters. The first and fourth terms describe the magnitude and the trend of the $\alpha$ preformation probability, the second and third terms show a parabolic dependence of $log_{10}P_\alpha$ taking into account the valence protons (neutrons) and holes, the last term relates to the correlation of valence neutron-proton~\cite{Guo15}. 

\section{Numerical results and discussion}

In this work, we systematically study and analyse the $\alpha$ decay half-lives, including 270 odd-$A$ nuclei, i.e., even $Z$, odd $N$ (even-odd) and odd $Z$, even $N$ (odd-even) nuclei. The experimental data of $\alpha$ decay energy and half-lives are calculated based on the NUBASE2012~\cite{Aud12}. We have previously preformed the systematic study for even-even nuclei with isospin dependent nuclear potential and a description of the $\alpha$ preformation probability~\cite{Sun161}. Compared to the even-even nuclei, the unpaired proton or neutron in odd-$A$ nuclei results in complex single particle level structure and even unfavored $\alpha$ decay. For unfavored transition, the additional centrifugal potential will increases the height of total potential barrier and decreases the penetrate probability. Besides, it is awkward that the odd nucleon may hinder the cluster behaviour of nucleon pairs by the Pauli blocking effect~\cite{Xu99}. Because of these difficulties, calculations on $\alpha$ decay half-lives for odd-$A$ nuclei deserve further study and improvement. There is no doubt that the systematic study will help to design experiments and test the validity of new measured data in future.

\begin{figure}
\resizebox{0.5\textwidth}{!}{%
  \includegraphics{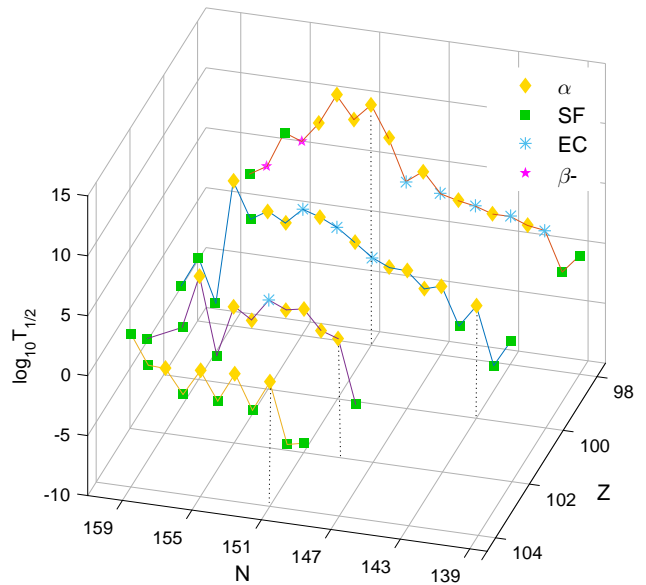}
}
\caption{Measured half-lives of ground states of Cf, Fm, No and Rf isotopes. Main decay modes are indicated by different symbols given in the legend.}
\label{fig1}
\end{figure}

At first, we discuss the odd-even staggering effect of $\alpha$ decay half-lives. The measured half-lives of ground states of Cf, Fm, No and Rf isotopes are plotted in Fig. 1. Using yellow diamonds, green squares, blue asterisks and peach stars denote the main decay modes, i.e., $\alpha$ decay, spontaneous fission (SF), electron capture (EC) and $\beta$- decay, respectively. From the picture we can clearly see that from Cf to Rf isotopes the average half-life drops off with the increasing of proton number. Moreover, the odd-$A$ nuclei of Cf isotopes tend to transform through EC and $\beta$- decay because their $\alpha$ decay half-lives are longer than the corresponding $\beta$ decay ones~\cite{Ren14}. As shown in Fig. 1, the half-lives of $^{249, 251}$Cf isotopes are longer than the even-even isotopes, while the half-lives of EC nuclei $^{239, 241, 243, 245, 247}$Cf and $\beta$- nuclei $^{253, 255}$Cf isotopes are as much as or even shorter than their neighboring ones. Additionally, the odd-$A$ nuclei $^{243, 257}$Fm, $^{259}$No and $^{255, 257, 259, 261}$Rf isotopes prefer $\alpha$ decay but not SF as the main decay mode, which are not similar to their neighboring nuclei, indicating that the odd nucleon blocking effect seems play a more important role on SF than on $\alpha$ decay.

\begin{figure}
\resizebox{0.5\textwidth}{!}{%
  \includegraphics{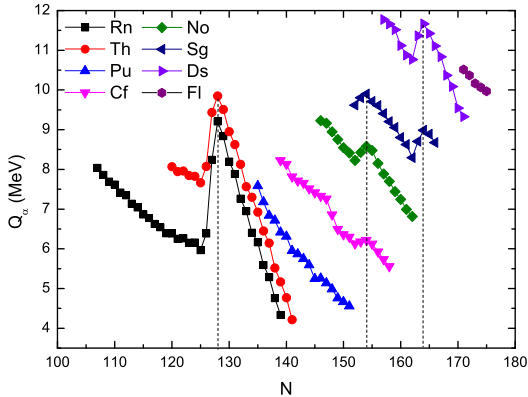}
}
\caption{The $\alpha$ decay energy between ground states of the parent and daughter nuclei of Rn, Th, Pu, Cf, No, Sg and Ds isotopes.}
\label{fig2}
\end{figure}

The $\alpha$ decay energy is one of the most important inputs for calculation of the half-life without obvious odd-even staggering effect. $\alpha$ decay was first successfully explained by Gamow~\cite{Gam28} and independently by Gurney and Condon~\cite{Gur28} as the penetration through the Coulomb barrier, leading to the $Q_\alpha^{-1/2}$ dependence of the Geiger-Nuttall law. Besides, the $\alpha$ preformation probability also depend linearly on $Q_\alpha^{-1/2}$, especially for nuclei with neutron number below $N$=126~\cite{Qi14}. The $\alpha$ decay energies which are obtained by the atomic mass difference between the ground states of the parent and daughter nuclei, including Rn, Th, Pu, Cf, No, Sg and Ds isotopes, are shown in Fig. 2. It can be seen that $\alpha$ decay energy approximates several or over ten MeVs, and decreases with increasing neutron number on an isotope chain without obvious odd-even staggering effect. The shell closure will abruptly increases $\alpha$ decay energy, which takes local maximum value at the place of magic number plus two additional nucleons as shown by the dotted lines at $N$=128, 154 and 164. It should be special mentioned the strong dependence of $\alpha$ decay half-life on decay energy, and the decay energy is relative large resulting in the unstability of superheavy nuclei. In view of the general downward trend for $\alpha$ decay energy with increasing of neutron number, new synthesised more neutron-rich superheavy nuclei would survive longer. 

Now that the $\alpha$ decay energy has no obvious odd-even differences, the odd-even staggering effect of $\alpha$ decay half-lives maybe comes from the $\alpha$ particle preformation probability mostly. According to the method of treating even-even nuclei, we extract the $\alpha$ particle preformation probabilities from ratios of the calculated $\alpha$ decay half-lives within the two-potential approach to the experimental data by eq. (9). Then the trend of the extracted $\alpha$ preformation probabilities can be described by eq. (10). Using the esitimated probabilities, the calculated half-lives can well reproduce the experimental data~\cite{Sun161}. For odd-$A$ nuclei, the unpaired nucleon is easy to be scattered to the higher single particle level. It indicates that the $\alpha$ decay daughter nuclei have uncertain configuration, resulting in unfavored $\alpha$ decay is quite common. In the following text, we will systematically study the $\alpha$ decay half-lives for odd-$A$ nuclei.

The whole $\alpha$ decay nuclei is divided into 4 regions according to the major shell closures~\cite{Zha08,Sun161}. In the chart of nuclides for $\alpha$ decay, there are 4 regions, region I: 50$<Z<$82 and 82$<N<$126, region II: 82$<Z<$126 and 82$<N<$126, region III: 82$<Z<$126 and 126$<N<$152, region IV: 82$<Z<$126 and 152$<N<$184. Here $Z$=50, 82 and $N$=82, 126 are the well-known spherical shell closures, while $N$=152 is the deformed neutron shell given by many experimental data, and $Z$=126, $N$=184 are the hypothetical spherical closed shells predicted by different theoritical models. The parameters of $\alpha$ preformation probaility of eq. (10) and the calculated $\alpha$ decay half-lives in different regions are listed in Tables from I to V, respectively. The nuclei in region III are separately treated for the cases of even-odd and odd-even nuclei as shown in Table I.  

\begin{table}
\caption{The parameters of $\alpha$ preformation probaility of eq. (10) in different regions. III.1 and III.2 denote the results for even-odd and odd-even nuclei in region III, respectively.}
\label{tab1}       
\begin{tabular}{cccccc}
\hline\noalign{\smallskip}
Region & a & b & c & d & e \\
\noalign{\smallskip}\hline\noalign{\smallskip}
I & -1.8782 & 0.0013 & 0.0015 &	0.0049 & -0.0009 \\
II & -7.0561 & 0.004 & 0.0027 & 0.0248 & -0.0017 \\
III.1 & 51.2962 & 0.0044 & 0.0128 & -0.2517 & 0.0172 \\
III.2 & 10.6825 & 0.0042 & 0.0035 & -0.0587 & 0.0015 \\
IV & 5.3064 & -0.0085 & -0.003 & -0.009 & 0.0004 \\
\noalign{\smallskip}\hline
\end{tabular}
\end{table}

\begin{figure}
\centering
\subfigure{\includegraphics[width=0.5\textwidth]{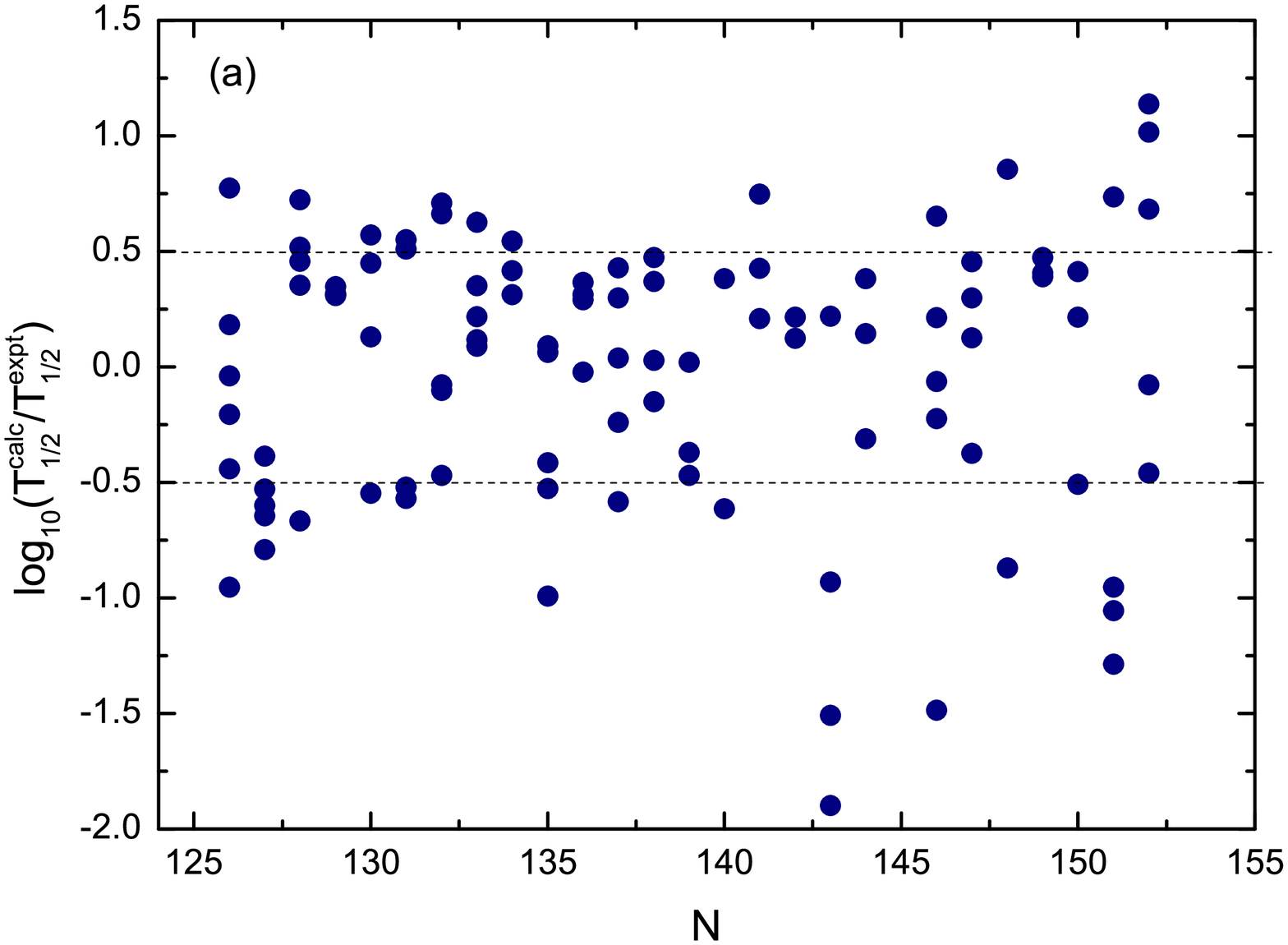}}
\subfigure{\includegraphics[width=0.5\textwidth]{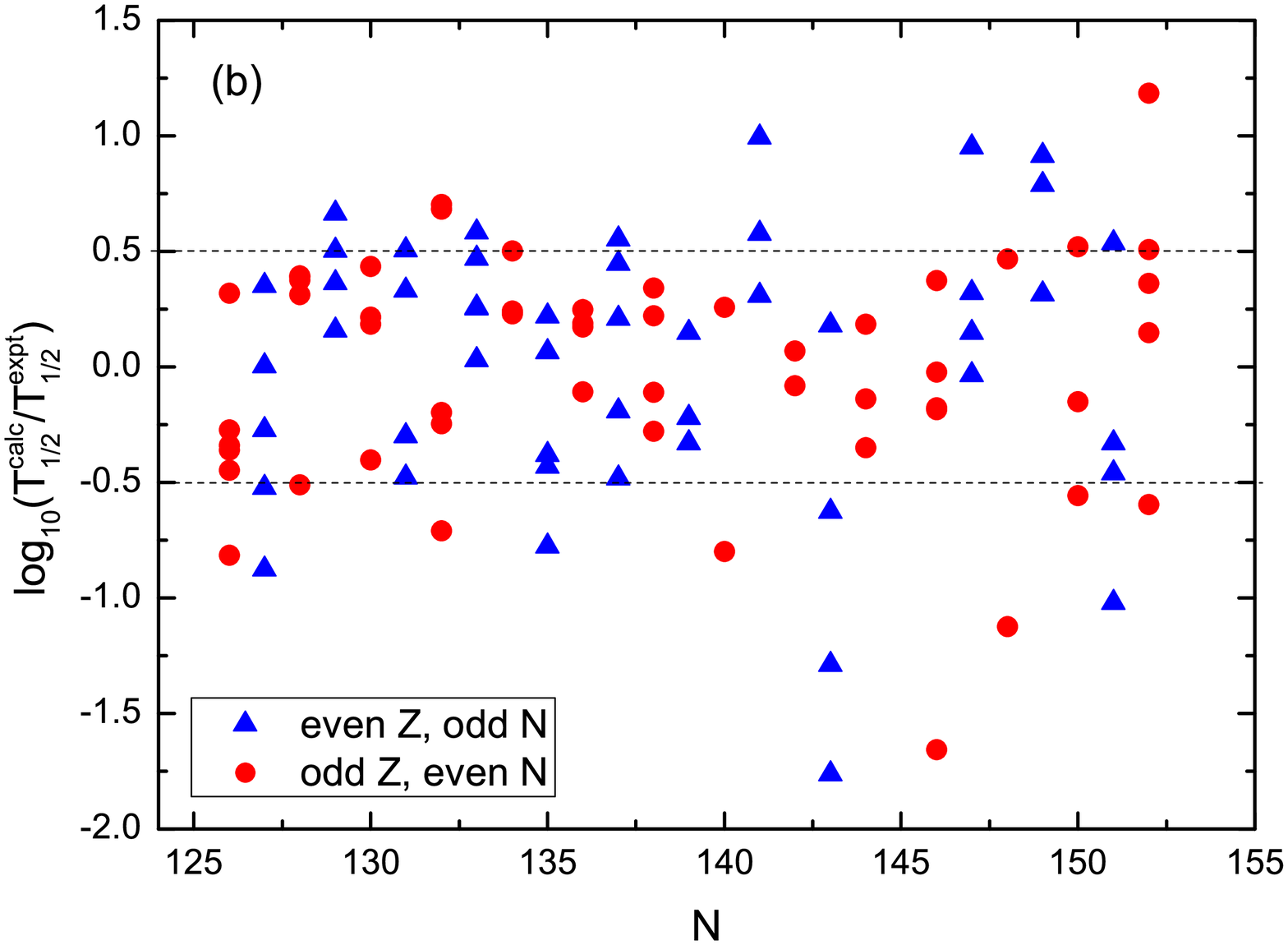}}
\caption{Logarithmic of deviations between the calculated half-lives and experimental data as a function of neutron numbers, showing the comparison of (a) in a unified way, (b) separately treating the $\alpha$ preformation probability of even $Z$, odd $N$ and odd $Z$, even $N$ nuclei in region III. }
\label{fig3}
\end{figure}

\begin{figure}
\resizebox{0.5\textwidth}{!}{%
  \includegraphics{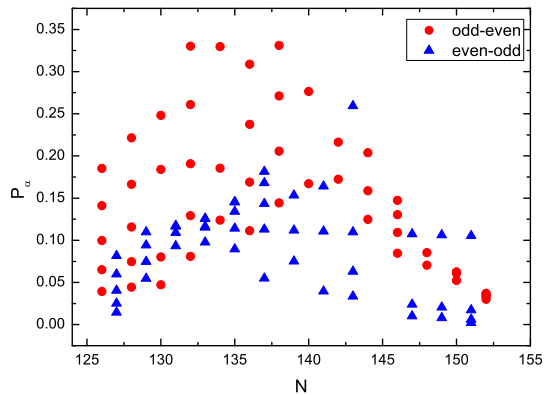}
}
\caption{The comparison of the estimated $\alpha$ preformation probability between the even-odd and odd-even nuclei in region III as a function of the neutron numbers of the parent nuclei.}
\label{fig4}
\end{figure}

The blocking effects coming from the unpaired proton and neutron are different, which could not be ignored especially for the nuclei in region III. The comparison between in a unified way and separately treating the $\alpha$ preformation probability of even-odd and odd-even nuclei in region III is shown in Fig. 3. We can see that in the case of separately treating, in Fig. 3(b), the deviations between the calculated half-livse and experimental data are smaller than those in a unified way in Fig. 3(a). And the estimated $\alpha$ preformation probability of even-odd and odd-even nuclei in region III are denoted by blue triangles and red circles in Fig. 4, respectively. In general, the $\alpha$ preformation probabilities of even-odd nuclei are more smaller than the odd-even ones, which is consistent with the results within the Viola-Seaborg formula~\cite{Don05}. 

\begin{figure}
\resizebox{0.5\textwidth}{!}{%
  \includegraphics{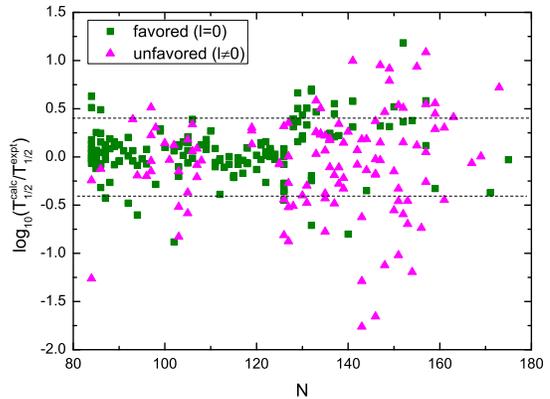}
}
\caption{Logarithmic of deviations between the calculated half-lives and experimental data as a function of neutron numbers of the parent nuclei.}
\label{fig5}
\end{figure}

From Table II to V, we show the experimental $\alpha$ decay half-lives data and numerical results within the two-potential approach of odd-$A$ nuclei. For all the tables, the first column denotes the $\alpha$ transition from the parent to daughter nuclei, including ground and isomeric states. The next four columns express the spin-parity transition, $\alpha$ decay energy, the minimal angular momentum quantum number carried out by the emitted $\alpha$ particle and the experimental $\alpha$ decay half-lives, respectively. The sixth and seventh columns denote the calculated $\alpha$ decay half-lives and $\alpha$ preformation probabilities, respectively. To clearly show the results, we plot the deviations between the calculated half-lives and experimental data in Fig. 5. The green squares and magenta triangles denote the favored and unfavored $\alpha$ decay, respectively. For nuclei with neutron number below $N$=126 in regions I and II, the results can better reproduce the experimental half-lives than those in regions III and IV where the unfavored $\alpha$ decay with nonzero angular momentum transfer is quite common. And in the case of unfavored $\alpha$ decay, the $\alpha$ preformation probability seems to be smaller and uncertain, which is worthy to further study in the future. 

\section{Summary}

In summary, we systematically study the $\alpha$ decay half-lives for odd-$A$ nuclei within the two-potential approach. The $\alpha$ preformation probabilities are extracted from the measured $\alpha$ decay half-lives, and estimated with the analytic formula taking into account the shell effect and proton-neutron correlation. The odd-even staggering effect of half-lives is investigated. We found that electron capture and $\beta$- decay are insensitive to unpaired nucleon, and the odd nucleon blocking effect play a more important role on spontaneous fission than $\alpha$ decay. The fact that $\alpha$ decay energy without obvious odd-even effect results in the $\alpha$ preformation probability are directly related to the unpaired nucleon. The difference of blocking effect between odd proton and neutron can not be ignored, and the $\alpha$ preformation probabilities of even-$Z$, odd-$N$ nuclei are smaller than the odd-$Z$, even-$N$ ones. In general, the numerical results of $\alpha$ decay half-lives can well reproduce the experimental data for odd-$A$ nuclei just as the case of even-even nuclei, and the unfavored $\alpha$ decay deserves further study. 

\begin{acknowledgments}
This work is supported in part by the National Natural Science Foundation of China (Grant No. 11205083), the construct program of the key discipline in hunan province, the Research Foundation of Education Bureau of Hunan Province, China (Grant No. 15A159), the Natural Science Foundation of Hunan Province, China (Grant No. 2015JJ3103, 2015JJ2123), the Innovation Group of Nuclear and Particle Physics in USC, Hunan Provincial Innovation Foundation For Postgraduate (Grant No. CX2015B398).
\end{acknowledgments}

%

\clearpage
\begingroup
\renewcommand*{\arraystretch}{1.3}
\begin{longtable*}{ccccccc}
\caption{Calculations of $\alpha$ decay half-lives of odd-$A$ nuclei in region I within a two-potential approach.}\label{tab2} \\
\hline 
$\alpha$ transition & $I_i^\pi \to I_j^\pi$ & $Q_\alpha$(MeV) &$l_\text{min}$ & $T^\text{expt}_{1/2}$(s) & $T^\text{calc}_{1/2}$(s) & $P_\alpha$ \\ \hline
\endfirsthead
\multicolumn{7}{c}%
{{\tablename\ \thetable{} -- continued}} \\
\hline 
$\alpha$ transition & $I_i^\pi \to I_j^\pi$ & $Q_\alpha$(MeV) &$l_\text{min}$ & $T^\text{expt}_{1/2}$(s) & $T^\text{calc}_{1/2}$(s) & $P_\alpha$ \\ \hline
\endhead
\hline \\
\endfoot
\hline \hline
\endlastfoot
$^{145}$Pm$\to$$^{141}$Pr&5/2$^+$$\to$5/2$^+$&2.324&0&1.99$\times10^{17}$&6.47$\times10^{17}$&0.175\\
$^{147}$Sm$\to$$^{143}$Nd&7/2$^-$$\to$7/2$^-$&2.311&0&3.36$\times10^{18}$&6.03$\times10^{18}$&0.205\\
$^{147}$Eu$\to$$^{143}$Pm&5/2$^+$$\to$5/2$^+$&2.991&0&9.46$\times10^{10}$&4.02$\times10^{11}$&0.187\\
$^{149}$Gd$\to$$^{145}$Sm&7/2$^-$$\to$7/2$^-$&3.099&0&1.86$\times10^{11}$&1.64$\times10^{11}$&0.214\\
$^{149}$Tb$\to$$^{145}$Eu&1/2$^+$$\to$5/2$^+$&4.078&2&8.88$\times10^{4}$&5.07$\times10^{4}$&0.194\\
$^{149}$Tb$^m$$\to$$^{145}$Eu&11/2$^-$$\to$5/2$^+$&4.114&3&1.13$\times10^{6}$&6.21$\times10^{4}$&0.194\\
$^{151}$Tb$\to$$^{147}$Eu&1/2$^{(+)}$$\to$5/2$^+$&3.496&2&6.67$\times10^{8}$&5.03$\times10^{8}$&0.242\\
$^{151}$Dy$\to$$^{147}$Gd&7/2$^{(-)}$$\to$7/2$^-$&4.18&0&1.92$\times10^{4}$&1.87$\times10^{4}$&0.219\\
$^{151}$Ho$\to$$^{147}$Tb$^m$&11/2$^{(-)}$$\to$11/2$^-\#$&4.644&0&1.6$\times10^{2}$&1.79$\times10^{2}$&0.197\\
$^{151}$Ho$^m$$\to$$^{147}$Tb&1/2$^{(+)}$$\to$(1/2$^+$)&4.736&0&6.13$\times10^{1}$&5.85$\times10^{1}$&0.197\\
$^{153}$Dy$\to$$^{149}$Gd&7/2$^{(-)}$$\to$7/2$^-$&3.559&0&2.45$\times10^{8}$&3.02$\times10^{8}$&0.268\\
$^{153}$Ho$\to$$^{149}$Tb$^m$&11/2$^-$$\to$11/2$^-$&4.016&0&2.36$\times10^{5}$&7.29$\times10^{5}$&0.243\\
$^{153}$Ho$^m$$\to$$^{149}$Tb&1/2$^+$$\to$1/2$^+$&4.121&0&3.1$\times10^{5}$&1.49$\times10^{5}$&0.243\\
$^{153}$Er$\to$$^{149}$Dy&7/2$^{(-)}$$\to$7/2$^{(-)}$&4.802&0&7$\times10^{1}$&7.94$\times10^{1}$&0.219\\
$^{153}$Tm$\to$$^{149}$Ho&(11/2$^-$)$\to$(11/2$^-$)&5.248&0&1.63$\times10^{0}$&2.18$\times10^{0}$&0.195\\
$^{153}$Tm$^m$$\to$$^{149}$Ho$^m$&(1/2$^+$)$\to$(1/2$^+$)&5.242&0&2.72$\times10^{0}$&2.33$\times10^{0}$&0.195\\
$^{155}$Er$\to$$^{151}$Dy&(7/2$^-$)$\to$7/2$^{(-)}$&4.118&0&1.45$\times10^{6}$&5.44$\times10^{5}$&0.265\\
$^{155}$Tm$\to$$^{151}$Ho&(11/2$^-$)$\to$11/2$^{(-)}$&4.572&0&2.43$\times10^{3}$&4.28$\times10^{3}$&0.239\\
$^{155}$Yb$\to$$^{151}$Er&(7/2$^-$)$\to$(7/2$^-$)&5.338&0&2.01$\times10^{0}$&2.36$\times10^{0}$&0.213\\
$^{155}$Lu$\to$$^{151}$Tm&(11/2$^-$)$\to$(11/2$^-$)&5.803&0&7.62$\times10^{-2}$&9.13$\times10^{-2}$&0.188\\
$^{155}$Lu$^m$$\to$$^{151}$Tm$^m$&(1/2$^+$)$\to$(1/2$^+$)&5.73&0&1.82$\times10^{-1}$&1.81$\times10^{-1}$&0.188\\
$^{157}$Yb$\to$$^{153}$Er&7/2$^-$$\to$7/2$^{(-)}$&4.621&0&7.72$\times10^{3}$&7.26$\times10^{3}$&0.256\\
$^{157}$Lu$^m$$\to$$^{153}$Tm&(11/2$^-$)$\to$(11/2$^-$)&5.128&0&7.98$\times10^{1}$&6.56$\times10^{1}$&0.229\\
$^{157}$Hf$\to$$^{153}$Yb&7/2$^-$$\to$7/2$^-\#$&5.885&0&1.34$\times10^{-1}$&1.12$\times10^{-1}$&0.202\\
$^{159}$Ta$\to$$^{155}$Lu$^m$&1/2$^+$$\to$(1/2$^+$)&5.66&0&3.06$\times10^{0}$&2.6$\times10^{0}$&0.214\\
$^{159}$Ta$^m$$\to$$^{155}$Lu&11/2$^-$$\to$(11/2$^-$)&5.744&0&1.02$\times10^{0}$&1.14$\times10^{0}$&0.214\\
$^{159}$W$\to$$^{155}$Hf&7/2$^-\#$$\to$7/2$^-\#$&6.445&0&1$\times10^{-2}$&7.58$\times10^{-3}$&0.187\\
$^{159}$Re$^m$$\to$$^{155}$Ta&11/2$^-$$\to$(11/2$^-$)&6.965&0&2.88$\times10^{-4}$&4.32$\times10^{-4}$&0.163\\
$^{161}$W$\to$$^{157}$Hf&7/2$^-\#$$\to$7/2$^-$&5.915&0&5.6$\times10^{-1}$&6.04$\times10^{-1}$&0.222\\
$^{161}$Re$^m$$\to$$^{157}$Ta$^m$&11/2$^-$$\to$11/2$^-$&6.425&0&1.58$\times10^{-2}$&2.26$\times10^{-2}$&0.195\\
$^{161}$Os$\to$$^{157}$W&(7/2$^-$)$\to$(7/2$^-$)&7.065&0&6.4$\times10^{-4}$&4.89$\times10^{-4}$&0.169\\
$^{163}$W$\to$$^{159}$Hf&7/2$^-$$\to$7/2$^-$&5.518&0&1.88$\times10^{1}$&2.41$\times10^{1}$&0.255\\
$^{163}$Re$\to$$^{159}$Ta&1/2$^+$$\to$1/2$^+$&6.012&0&1.22$\times10^{0}$&6.64$\times10^{-1}$&0.226\\
$^{163}$Re$^m$$\to$$^{159}$Ta$^m$&11/2$^-$$\to$11/2$^-$&6.068&0&3.24$\times10^{-1}$&3.95$\times10^{-1}$&0.226\\
$^{163}$Os$\to$$^{159}$W&7/2$^-\#$$\to$7/2$^-\#$&6.675&0&5.5$\times10^{-3}$&7.38$\times10^{-3}$&0.199\\
$^{165}$Re$^m$$\to$$^{161}$Ta$^m$&(11/2$^-$)$\to$(11/2$^-$)&5.66&0&1.78$\times10^{1}$&1.62$\times10^{1}$&0.256\\
$^{165}$Os$\to$$^{161}$W&(7/2$^-$)$\to$7/2$^-\#$&6.335&0&1.18$\times10^{-1}$&9.8$\times10^{-2}$&0.227\\
$^{165}$Ir$^m$$\to$$^{161}$Re$^m$&(11/2$^-$)$\to$11/2$^-$&6.885&0&2.31$\times10^{-3}$&3.54$\times10^{-3}$&0.2\\
$^{167}$Re$^m$$\to$$^{163}$Ta&1/2$^+$$\to$1/2$^+$&5.405&0&5.9$\times10^{2}$&1.96$\times10^{2}$&0.282\\
$^{167}$Os$\to$$^{163}$W&7/2$^-$$\to$7/2$^-$&5.985&0&1.65$\times10^{0}$&1.9$\times10^{0}$&0.252\\
$^{167}$Ir$\to$$^{163}$Re&1/2$^+$$\to$1/2$^+$&6.504&0&6.81$\times10^{-2}$&6.15$\times10^{-2}$&0.224\\
$^{167}$Ir$^m$$\to$$^{163}$Re$^m$&11/2$^-$$\to$11/2$^-$&6.561&0&2.86$\times10^{-2}$&3.8$\times10^{-2}$&0.224\\
$^{167}$Pt$\to$$^{163}$Os&7/2$^-\#$$\to$7/2$^-\#$&7.155&0&8$\times10^{-4}$&1.15$\times10^{-3}$&0.197\\
$^{169}$Re$^m$$\to$$^{165}$Ta&(1/2$^+$,3/2$^+$)$\to$(1/2$^+$,3/2$^+$)&5.189&0&7.55$\times10^{3}$&1.89$\times10^{3}$&0.301\\
$^{169}$Os$\to$$^{165}$W&(5/2$^-$)$\to$(5/2$^-$)&5.713&0&2.53$\times10^{1}$&2.29$\times10^{1}$&0.272\\
$^{169}$Ir$\to$$^{165}$Re&(1/2$^+$)$\to$(1/2$^+$)&6.141&0&7.84$\times10^{-1}$&1.25$\times10^{0}$&0.244\\
$^{169}$Ir$^m$$\to$$^{165}$Re$^m$&(11/2$^-$)$\to$(11/2$^-$)&6.266&0&3.9$\times10^{-1}$&4$\times10^{-1}$&0.244\\
$^{171}$Ir$^m$$\to$$^{167}$Re&(11/2$^-$)$\to$(9/2$^-$)&6.155&2&2.72$\times10^{0}$&1.74$\times10^{0}$&0.259\\
$^{171}$Pt$\to$$^{167}$Os&7/2$^-$$\to$7/2$^-$&6.605&0&5.06$\times10^{-2}$&5.86$\times10^{-2}$&0.232\\
$^{173}$Ir$^m$$\to$$^{169}$Re&(11/2$^-$)$\to$(9/2$^-$)&5.942&2&1.83$\times10^{1}$&1.16$\times10^{1}$&0.268\\
$^{173}$Pt$\to$$^{169}$Os&(5/2$^-$)$\to$(5/2$^-$)&6.358&0&4.44$\times10^{-1}$&4.23$\times10^{-1}$&0.242\\
$^{173}$Au$\to$$^{169}$Ir&(1/2$^+$)$\to$(1/2$^+$)&6.837&0&2.91$\times10^{-2}$&2.38$\times10^{-2}$&0.217\\
$^{173}$Hg$\to$$^{169}$Pt&3/2$^-\#$$\to$(7/2$^-$)&7.375&2&9.1$\times10^{-4}$&2.23$\times10^{-3}$&0.193\\
$^{175}$Ir$\to$$^{171}$Re&5/2$^-\#$$\to$(9/2$^-$)&5.431&2&1.06$\times10^{3}$&2.13$\times10^{3}$&0.269\\
$^{175}$Pt$\to$$^{171}$Os&(7/2$^-$)$\to$(5/2$^-$)&6.178&2&3.95$\times10^{0}$&3.59$\times10^{0}$&0.246\\
$^{175}$Au$\to$$^{171}$Ir&1/2$^+$$\to$1/2$^+$&6.575&0&2.16$\times10^{-1}$&1.83$\times10^{-1}$&0.223\\
$^{177}$Ir$\to$$^{173}$Re&5/2$^-$$\to$(5/2$^-$)&5.082&0&5$\times10^{4}$&6.3$\times10^{4}$&0.263\\
$^{177}$Pt$\to$$^{173}$Os&5/2$^-$$\to$(5/2$^-$)&5.643&0&1.86$\times10^{2}$&3.63$\times10^{2}$&0.243\\
$^{177}$Au$\to$$^{173}$Ir&(1/2$^+$,3/2$^+$)$\to$(1/2$^+$,3/2$^+$)&6.298&0&3.65$\times10^{0}$&1.9$\times10^{0}$&0.222\\
$^{177}$Hg$\to$$^{173}$Pt&(7/2$^-$)$\to$(5/2$^-$)&6.735&2&1.5$\times10^{-1}$&2.5$\times10^{-1}$&0.201\\
$^{177}$Tl$\to$$^{173}$Au&(1/2$^+$)$\to$(1/2$^+$)&7.066&0&2.47$\times10^{-2}$&2.96$\times10^{-2}$&0.181\\
$^{179}$Pt$\to$$^{175}$Os&1/2$^-$$\to$(5/2$^-$)&5.412&2&8.83$\times10^{3}$&8.23$\times10^{3}$&0.233\\
$^{179}$Au$\to$$^{175}$Ir&(1/2$^+$,3/2$^+$)$\to$5/2$^-\#$&5.98&1&3.23$\times10^{1}$&4.48$\times10^{1}$&0.215\\
$^{179}$Hg$\to$$^{175}$Pt&7/2$^-$$\to$(7/2$^-$)&6.351&0&1.91$\times10^{0}$&3.55$\times10^{0}$&0.197\\
$^{179}$Pb$\to$$^{175}$Hg&(9/2$^-$)$\to$(7/2$^-$)&7.595&2&3.9$\times10^{-3}$&2.76$\times10^{-3}$&0.161\\
$^{181}$Pt$\to$$^{177}$Os&1/2$^-$$\to$1/2$^-$&5.15&0&7.03$\times10^{4}$&9.38$\times10^{4}$&0.218\\
$^{181}$Au$\to$$^{177}$Ir&(3/2$^-$)$\to$5/2$^-$&5.751&2&5.07$\times10^{2}$&6.73$\times10^{2}$&0.203\\
$^{183}$Pt$\to$$^{179}$Os&1/2$^-$$\to$1/2$^-$&4.822&0&4.06$\times10^{6}$&6.54$\times10^{6}$&0.198\\
$^{183}$Au$\to$$^{179}$Ir&5/2$^-$$\to$(5/2)$^-$&5.465&0&7.78$\times10^{3}$&8.07$\times10^{3}$&0.187\\
$^{183}$Hg$\to$$^{179}$Pt&1/2$^-$$\to$1/2$^-$&6.038&0&8.03$\times10^{1}$&6.33$\times10^{1}$&0.174\\
$^{183}$Tl$\to$$^{179}$Au&1/2$^{(+)}$$\to$(1/2$^+$,3/2$^+$)&5.977&0&3.45$\times10^{2}$&3.99$\times10^{2}$&0.161\\
$^{185}$Au$\to$$^{181}$Ir&5/2$^-$$\to$5/2$^-$&5.18&0&9.81$\times10^{4}$&2.41$\times10^{5}$&0.167\\
$^{185}$Hg$\to$$^{181}$Pt&1/2$^-$$\to$1/2$^-$&5.773&0&8.18$\times10^{2}$&9.56$\times10^{2}$&0.157\\
$^{185}$Pb$\to$$^{181}$Hg&3/2$^-$$\to$1/2$^{(-\#)}$&6.695&2&1.85$\times10^{1}$&2.74$\times10^{0}$&0.136\\
$^{185}$Pb$^m$$\to$$^{181}$Hg$^m$&13/2$^+$$\to$13/2$^+$&6.555&0&8.14$\times10^{0}$&5.13$\times10^{0}$&0.136\\
$^{187}$Tl$^m$$\to$$^{183}$Au&(9/2$^-$)$\to$5/2$^-$&5.656&2&1.04$\times10^{4}$&2.26$\times10^{4}$&0.13\\
$^{187}$Pb$\to$$^{183}$Hg&3/2$^-$$\to$1/2$^-$&6.393&2&1.6$\times10^{2}$&4.16$\times10^{1}$&0.121\\
$^{187}$Pb$^m$$\to$$^{183}$Hg$^m$&13/2$^+$$\to$13/2$^+\#$&6.208&0&1.53$\times10^{2}$&1.32$\times10^{2}$&0.121\\
$^{189}$Pb$\to$$^{185}$Hg&3/2$^-$$\to$1/2$^-$&5.871&2&1.26$\times10^{4}$&7.77$\times10^{3}$&0.106\\
$^{191}$Pb$^m$$\to$$^{187}$Hg$^m$&13/2$^{(+)}$$\to$13/2$^+$&5.404&0&6.54$\times10^{5}$&8.88$\times10^{5}$&0.089\\
\end{longtable*}
\endgroup

\clearpage
\begingroup
\renewcommand*{\arraystretch}{1.3}
\begin{longtable*}{ccccccc}
\caption{Same as Table II, but for odd-$A$ nuclei in region II.}\label{tab3} \\
\hline 
$\alpha$ transition & $I_i^\pi \to I_j^\pi$ & $Q_\alpha$(MeV) &$l_\text{min}$ & $T^\text{expt}_{1/2}$(s) & $T^\text{calc}_{1/2}$(s) & $P_\alpha$ \\ \hline
\endfirsthead
\multicolumn{7}{c}%
{{\tablename\ \thetable{} -- continued}} \\
\hline 
$\alpha$ transition & $I_i^\pi \to I_j^\pi$ & $Q_\alpha$(MeV) &$l_\text{min}$ & $T^\text{expt}_{1/2}$(s) & $T^\text{calc}_{1/2}$(s) & $P_\alpha$ \\ \hline
\endhead
\hline \\
\endfoot
\hline \hline
\endlastfoot
$^{179}$Pb$\to$$^{175}$Hg&(9/2$^-$)$\to$(7/2$^-$)&7.595&2&3.9$\times10^{-3}$&1.27$\times10^{-2}$&0.034\\
$^{185}$Pb$\to$$^{181}$Hg&3/2$^-$$\to$1/2$^{(-\#)}$&6.695&2&1.85$\times10^{1}$&5.62$\times10^{0}$&0.065\\
$^{185}$Pb$^m$$\to$$^{181}$Hg$^m$&13/2$^+$$\to$13/2$^+$&6.555&0&8.14$\times10^{0}$&1.05$\times10^{1}$&0.065\\
$^{185}$Bi$^m$$\to$$^{181}$Tl&1/2$^+$$\to$1/2$^+$&8.218&0&5.8$\times10^{-4}$&7.62$\times10^{-5}$&0.087\\
$^{187}$Pb$\to$$^{183}$Hg&3/2$^-$$\to$1/2$^-$&6.393&2&1.6$\times10^{2}$&6.82$\times10^{1}$&0.072\\
$^{187}$Pb$^m$$\to$$^{183}$Hg$^m$&13/2$^+$$\to$13/2$^+\#$&6.208&0&1.53$\times10^{2}$&2.16$\times10^{2}$&0.072\\
$^{187}$Bi$^m$$\to$$^{183}$Tl&1/2$^+\#$$\to$1/2$^{(+)}$&7.887&0&3.7$\times10^{-4}$&5.27$\times10^{-4}$&0.099\\
$^{187}$Po$\to$$^{183}$Pb&(1/2$^-$,5/2$^-$)$\to$3/2$^-$&7.976&2&1.4$\times10^{-3}$&9.91$\times10^{-4}$&0.133\\
$^{189}$Pb$\to$$^{185}$Hg&3/2$^-$$\to$1/2$^-$&5.871&2&1.26$\times10^{4}$&1.04$\times10^{4}$&0.077\\
$^{189}$Bi$^m$$\to$$^{185}$Tl&(1/2$^+$)$\to$1/2$^+\#$&7.452&0&9.8$\times10^{-3}$&9.55$\times10^{-3}$&0.108\\
$^{189}$Po$\to$$^{185}$Pb&(5/2$^-$)$\to$3/2$^-$&7.694&2&3.8$\times10^{-3}$&5.73$\times10^{-3}$&0.147\\
$^{191}$Pb$^m$$\to$$^{187}$Hg$^m$&13/2$^{(+)}$$\to$13/2$^+$&5.404&0&6.54$\times10^{5}$&9.93$\times10^{5}$&0.079\\
$^{191}$Bi$^m$$\to$$^{187}$Tl&(1/2$^+$)$\to$(1/2$^+$)&7.018&0&1.82$\times10^{-1}$&2.42$\times10^{-1}$&0.111\\
$^{191}$At$^m$$\to$$^{187}$Bi&(7/2$^-$)$\to$9/2$^-\#$&7.88&2&2.2$\times10^{-3}$&2.52$\times10^{-3}$&0.209\\
$^{193}$Bi$^m$$\to$$^{189}$Tl&(1/2$^+$)$\to$(1/2$^+$)&6.613&0&3.81$\times10^{0}$&7.05$\times10^{0}$&0.11\\
$^{193}$At$^m$$\to$$^{189}$Bi&7/2$^-\#$$\to$(9/2$^-$)&7.581&2&2.1$\times10^{-2}$&1.92$\times10^{-2}$&0.213\\
$^{193}$Rn$\to$$^{189}$Po&3/2$^-\#$$\to$(5/2$^-$)&8.04&2&1.15$\times10^{-3}$&1.4$\times10^{-3}$&0.287\\
$^{195}$Bi$^m$$\to$$^{191}$Tl&(1/2$^+$)$\to$(1/2$^+$)&6.232&0&2.64$\times10^{2}$&2.42$\times10^{2}$&0.103\\
$^{195}$Po$\to$$^{191}$Pb&(3/2$^-$)$\to$(3/2$^-$)&6.755&0&4.94$\times10^{0}$&3.91$\times10^{0}$&0.147\\
$^{195}$Po$^m$$\to$$^{191}$Pb$^m$&(13/2$^+$)$\to$13/2$^{(+)}$&6.84&0&2.13$\times10^{0}$&1.87$\times10^{0}$&0.147\\
$^{195}$Rn$\to$$^{191}$Po&(3/2$^-$)$\to$(3/2$^-$)&7.694&0&7$\times10^{-3}$&8.4$\times10^{-3}$&0.283\\
$^{195}$Rn$^m$$\to$$^{191}$Po$^m$&(13/2$^+$)$\to$(13/2$^+$)&7.714&0&6$\times10^{-3}$&7.21$\times10^{-3}$&0.283\\
$^{197}$Po$\to$$^{193}$Pb&(3/2$^-$)$\to$(3/2$^-$)&6.405&0&1.22$\times10^{2}$&9.54$\times10^{1}$&0.134\\
$^{197}$At$\to$$^{193}$Bi&(9/2$^-$)$\to$(9/2$^-$)&7.108&0&4.04$\times10^{-1}$&3.88$\times10^{-1}$&0.19\\
$^{197}$Rn$\to$$^{193}$Po&(3/2$^-$)$\to$(3/2$^-$)&7.415&0&5.4$\times10^{-2}$&6.6$\times10^{-2}$&0.265\\
$^{199}$Po$\to$$^{195}$Pb&3/2$^-\#$$\to$3/2$^-\#$&6.074&0&4.38$\times10^{3}$&2.67$\times10^{3}$&0.116\\
$^{199}$At$\to$$^{195}$Bi&(9/2$^-$)$\to$(9/2$^-$)&6.778&0&7.89$\times10^{0}$&6.63$\times10^{0}$&0.167\\
$^{199}$Fr$\to$$^{195}$At&1/2$^+\#$$\to$1/2$^+$&7.811&0&1.6$\times10^{-2}$&6.55$\times10^{-3}$&0.328\\
$^{201}$Po$\to$$^{197}$Pb&3/2$^-$$\to$3/2$^-$&5.799&0&8.28$\times10^{4}$&5.63$\times10^{4}$&0.095\\
$^{201}$At$\to$$^{197}$Bi&(9/2$^-$)$\to$(9/2$^-$)&6.473&0&1.2$\times10^{2}$&1.18$\times10^{2}$&0.14\\
$^{201}$Fr$\to$$^{197}$At&(9/2$^-$)$\to$(9/2$^-$)&7.515&0&6.2$\times10^{-2}$&6.25$\times10^{-2}$&0.283\\
$^{203}$Po$\to$$^{199}$Pb&5/2$^-$$\to$3/2$^-$&5.496&2&2$\times10^{6}$&4.02$\times10^{6}$&0.075\\
$^{203}$At$\to$$^{199}$Bi&9/2$^-$$\to$9/2$^-$&6.209&0&1.43$\times10^{3}$&1.77$\times10^{3}$&0.112\\
$^{203}$Rn$\to$$^{199}$Po&3/2$^-\#$$\to$3/2$^-\#$&6.63&0&6.67$\times10^{1}$&6.16$\times10^{1}$&0.163\\
$^{203}$Fr$\to$$^{199}$At&9/2$^-\#$$\to$(9/2$^-$)&7.274&0&5.5$\times10^{-1}$&4.61$\times10^{-1}$&0.233\\
$^{203}$Ra$\to$$^{199}$Rn&(3/2$^-$)$\to$(3/2$^-$)&7.745&0&3.6$\times10^{-2}$&2.28$\times10^{-2}$&0.326\\
$^{205}$Po$\to$$^{201}$Pb&5/2$^-$$\to$5/2$^-$&5.325&0&1.57$\times10^{7}$&2.21$\times10^{7}$&0.056\\
$^{205}$At$\to$$^{201}$Bi&9/2$^-$$\to$9/2$^-$&6.02&0&2.03$\times10^{4}$&1.51$\times10^{4}$&0.085\\
$^{205}$Rn$\to$$^{201}$Po&5/2$^-$$\to$3/2$^-$&6.39&2&6.9$\times10^{2}$&1.29$\times10^{3}$&0.126\\
$^{205}$Fr$\to$$^{201}$At&(9/2$^-$)$\to$(9/2$^-$)&7.054&0&3.82$\times10^{0}$&3.3$\times10^{0}$&0.183\\
$^{207}$Po$\to$$^{203}$Pb&5/2$^-$$\to$5/2$^-$&5.216&0&9.94$\times10^{7}$&1.15$\times10^{8}$&0.04\\
$^{207}$At$\to$$^{203}$Bi&9/2$^-$$\to$9/2$^-$&5.872&0&6.52$\times10^{4}$&9.38$\times10^{4}$&0.061\\
$^{207}$Rn$\to$$^{203}$Po&5/2$^-$$\to$5/2$^-$&6.251&0&2.64$\times10^{3}$&3.52$\times10^{3}$&0.093\\
$^{207}$Fr$\to$$^{203}$At&9/2$^-$$\to$9/2$^-$&6.894&0&1.56$\times10^{1}$&1.62$\times10^{1}$&0.137\\
$^{207}$Ra$\to$$^{203}$Rn&5/2$^-\#$$\to$3/2$^-\#$&7.274&2&1.6$\times10^{0}$&2.17$\times10^{0}$&0.198\\
$^{207}$Ac$\to$$^{203}$Fr&9/2$^-\#$$\to$9/2$^-\#$&7.849&0&3.1$\times10^{-2}$&2.64$\times10^{-2}$&0.279\\
$^{209}$Bi$\to$$^{205}$Tl&9/2$^-$$\to$1/2$^+$&3.137&5&6.28$\times10^{26}$&2.26$\times10^{26}$&0.017\\
$^{209}$At$\to$$^{205}$Bi&9/2$^-$$\to$9/2$^-$&5.757&0&4.75$\times10^{5}$&4.51$\times10^{5}$&0.042\\
$^{209}$Rn$\to$$^{205}$Po&5/2$^-$$\to$5/2$^-$&6.155&0&1.01$\times10^{4}$&1.23$\times10^{4}$&0.065\\
$^{209}$Fr$\to$$^{205}$At&9/2$^-$$\to$9/2$^-$&6.777&0&5.62$\times10^{1}$&5.83$\times10^{1}$&0.097\\
$^{209}$Ra$\to$$^{205}$Rn&5/2$^-$$\to$5/2$^-$&7.135&0&5.23$\times10^{0}$&4.92$\times10^{0}$&0.143\\
$^{211}$At$\to$$^{207}$Bi&9/2$^-$$\to$9/2$^-$&5.982&0&6.21$\times10^{4}$&5.19$\times10^{4}$&0.028\\
$^{211}$Fr$\to$$^{207}$At&9/2$^-$$\to$9/2$^-$&6.662&0&2.14$\times10^{2}$&2.27$\times10^{2}$&0.066\\
$^{211}$Ra$\to$$^{207}$Rn&5/2$^{(-)}$$\to$5/2$^-$&7.042&0&1.42$\times10^{1}$&1.43$\times10^{1}$&0.098\\
$^{211}$Ac$\to$$^{207}$Fr&9/2$^-\#$$\to$9/2$^-$&7.619&0&2.13$\times10^{-1}$&2.41$\times10^{-1}$&0.144\\
$^{213}$Fr$\to$$^{209}$At&9/2$^-$$\to$9/2$^-$&6.904&0&3.5$\times10^{1}$&3.56$\times10^{1}$&0.043\\
$^{213}$Ra$\to$$^{209}$Rn&1/2$^-$$\to$5/2$^-$&6.862&2&2.05$\times10^{2}$&1.72$\times10^{2}$&0.065\\
$^{213}$Pa$\to$$^{209}$Ac&9/2$^-\#$$\to$(9/2$^-$)&8.395&0&7$\times10^{-3}$&3.81$\times10^{-3}$&0.197\\
$^{215}$Ac$\to$$^{211}$Fr&9/2$^-$$\to$9/2$^-$&7.746&0&1.7$\times10^{-1}$&1.78$\times10^{-1}$&0.061\\
$^{215}$Th$\to$$^{211}$Ra&(1/2$^-$)$\to$5/2$^{(-)}$&7.665&2&1.2$\times10^{0}$&1.03$\times10^{0}$&0.09\\
$^{215}$Pa$\to$$^{211}$Ac&9/2$^-\#$$\to$9/2$^-\#$&8.245&0&1.4$\times10^{-2}$&1.49$\times10^{-2}$&0.13\\
$^{217}$Pa$\to$$^{213}$Ac&9/2$^-\#$$\to$9/2$^-\#$&8.485&0&3.48$\times10^{-3}$&4.21$\times10^{-3}$&0.081\\
\end{longtable*}
\endgroup

\clearpage
\begingroup
\renewcommand*{\arraystretch}{1.3}
\begin{longtable*}{ccccccc}
\caption{Same as Table II, but for odd-$A$ nuclei in region III.}\label{tab4} \\
\hline 
$\alpha$ transition & $I_i^\pi \to I_j^\pi$ & $Q_\alpha$(MeV) &$l_\text{min}$ & $T^\text{expt}_{1/2}$(s) & $T^\text{calc}_{1/2}$(s) & $P_\alpha$ \\ \hline
\endfirsthead
\multicolumn{7}{c}%
{{\tablename\ \thetable{} -- continued}} \\
\hline 
$\alpha$ transition & $I_i^\pi \to I_j^\pi$ & $Q_\alpha$(MeV) &$l_\text{min}$ & $T^\text{expt}_{1/2}$(s) & $T^\text{calc}_{1/2}$(s) & $P_\alpha$ \\ \hline
\endhead
\hline \\
\endfoot
\hline \hline
\endlastfoot
$^{209}$Bi$\to$$^{205}$Tl&9/2$^-$$\to$1/2$^+$&3.137&5&6.28$\times10^{26}$&9.64$\times10^{25}$&0.039\\
$^{211}$Bi$\to$$^{207}$Tl&9/2$^-$$\to$1/2$^+$&6.75&5&1.28$\times10^{2}$&3.96$\times10^{1}$&0.045\\
$^{211}$Po$\to$$^{207}$Pb&9/2$^+$$\to$1/2$^-$&7.594&5&5.16$\times10^{-1}$&6.88$\times10^{-2}$&0.082\\
$^{211}$At$\to$$^{207}$Bi&9/2$^-$$\to$9/2$^-$&5.982&0&6.21$\times10^{4}$&2.22$\times10^{4}$&0.065\\
$^{213}$Bi$\to$$^{209}$Tl&9/2$^-$$\to$(1/2$^+$)&5.983&5&1.31$\times10^{5}$&5.2$\times10^{4}$&0.047\\
$^{213}$Po$\to$$^{209}$Pb&9/2$^+$$\to$9/2$^+$&8.536&0&3.72$\times10^{-6}$&5.36$\times10^{-6}$&0.11\\
$^{213}$At$\to$$^{209}$Bi&9/2$^-$$\to$9/2$^-$&9.254&0&1.25$\times10^{-7}$&3.08$\times10^{-7}$&0.075\\
$^{213}$Rn$\to$$^{209}$Po&9/2$^+\#$$\to$1/2$^-$&8.243&5&1.95$\times10^{-2}$&5.85$\times10^{-3}$&0.06\\
$^{213}$Fr$\to$$^{209}$At&9/2$^-$$\to$9/2$^-$&6.904&0&3.5$\times10^{1}$&1.53$\times10^{1}$&0.1\\
$^{215}$Po$\to$$^{211}$Pb&9/2$^+$$\to$9/2$^+$&7.526&0&1.78$\times10^{-3}$&3.81$\times10^{-3}$&0.116\\
$^{215}$At$\to$$^{211}$Bi&9/2$^-$$\to$9/2$^-$&8.178&0&1$\times10^{-4}$&1.53$\times10^{-4}$&0.08\\
$^{215}$Rn$\to$$^{211}$Po&9/2$^+$$\to$9/2$^+$&8.839&0&2.3$\times10^{-6}$&5.31$\times10^{-6}$&0.094\\
$^{215}$Fr$\to$$^{211}$At&9/2$^-$$\to$9/2$^-$&9.54&0&8.6$\times10^{-8}$&2.04$\times10^{-7}$&0.116\\
$^{215}$Ra$\to$$^{211}$Rn&9/2$^+\#$$\to$1/2$^-$&8.864&5&1.67$\times10^{-3}$&8.93$\times10^{-4}$&0.04\\
$^{215}$Ac$\to$$^{211}$Fr&9/2$^-$$\to$9/2$^-$&7.746&0&1.7$\times10^{-1}$&7.76$\times10^{-2}$&0.141\\
$^{217}$Po$\to$$^{213}$Pb&(9/2$^+$)$\to$(9/2$^+$)&6.662&0&1.59$\times10^{0}$&4.68$\times10^{0}$&0.098\\
$^{217}$At$\to$$^{213}$Bi&9/2$^-$$\to$9/2$^-$&7.201&0&3.23$\times10^{-2}$&1.63$\times10^{-1}$&0.081\\
$^{217}$Rn$\to$$^{213}$Po&9/2$^+$$\to$9/2$^+$&7.887&0&5.4$\times10^{-4}$&1.73$\times10^{-3}$&0.117\\
$^{217}$Ra$\to$$^{213}$Rn&(9/2$^+$)$\to$9/2$^+\#$&9.161&0&1.63$\times10^{-6}$&5.2$\times10^{-6}$&0.075\\
$^{217}$Ac$\to$$^{213}$Fr&9/2$^-$$\to$9/2$^-$&9.832&0&6.9$\times10^{-8}$&1.42$\times10^{-7}$&0.166\\
$^{217}$Th$\to$$^{213}$Ra&9/2$^+\#$$\to$1/2$^-$&9.435&5&2.47$\times10^{-4}$&2.49$\times10^{-4}$&0.025\\
$^{217}$Pa$\to$$^{213}$Ac&9/2$^-\#$$\to$9/2$^-\#$&8.485&0&3.48$\times10^{-3}$&1.86$\times10^{-3}$&0.185\\
$^{217}$Pa$^m$$\to$$^{213}$Ac&29/2$^+\#$$\to$9/2$^-\#$&10.345&11&1.48$\times10^{-3}$&3.08$\times10^{-3}$&0.185\\
$^{219}$Rn$\to$$^{215}$Po&5/2$^+$$\to$9/2$^+$&6.946&2&3.96$\times10^{0}$&4.24$\times10^{0}$&0.115\\
$^{219}$Fr$\to$$^{215}$At&9/2$^-$$\to$9/2$^-$&7.449&0&2$\times10^{-2}$&9.62$\times10^{-2}$&0.129\\
$^{219}$Ra$\to$$^{215}$Rn&(7/2)$^+$$\to$9/2$^+$&8.138&2&1$\times10^{-2}$&3.32$\times10^{-3}$&0.109\\
$^{219}$Ac$\to$$^{215}$Fr&9/2$^-$$\to$9/2$^-$&8.827&0&1.18$\times10^{-5}$&3.22$\times10^{-5}$&0.184\\
$^{219}$Th$\to$$^{215}$Ra&9/2$^+\#$$\to$9/2$^+\#$&9.511&0&1.05$\times10^{-6}$&4.85$\times10^{-6}$&0.055\\
$^{219}$Pa$\to$$^{215}$Ac&9/2$^-$$\to$9/2$^-$&10.084&0&5.3$\times10^{-8}$&1.31$\times10^{-7}$&0.221\\
$^{219}$U$\to$$^{215}$Th&9/2$^+\#$$\to$(1/2$^-$)&9.943&5&5.5$\times10^{-5}$&1.24$\times10^{-4}$&0.015\\
$^{221}$Rn$\to$$^{217}$Po&7/2$^+$$\to$(9/2$^+$)&6.163&2&7.01$\times10^{3}$&8.13$\times10^{3}$&0.09\\
$^{221}$Fr$\to$$^{217}$At&5/2$^-$$\to$9/2$^-$&6.458&2&2.87$\times10^{2}$&9.14$\times10^{2}$&0.124\\
$^{221}$Ra$\to$$^{217}$Rn&5/2$^+$$\to$9/2$^+$&6.88&2&2.8$\times10^{1}$&5.07$\times10^{1}$&0.125\\
$^{221}$Ac$\to$$^{217}$Fr&9/2$^-\#$$\to$9/2$^-$&7.78&0&5.2$\times10^{-2}$&3.3$\times10^{-2}$&0.191\\
$^{221}$Th$\to$$^{217}$Ra&(7/2$^+$)$\to$(9/2$^+$)&8.625&2&1.68$\times10^{-3}$&8.46$\times10^{-4}$&0.093\\
$^{221}$Pa$\to$$^{217}$Ac&9/2$^-$$\to$9/2$^-$&9.251&0&5.9$\times10^{-6}$&9.67$\times10^{-6}$&0.248\\
$^{223}$Rn$\to$$^{219}$Po&7/2$^{(-\#)}$$\to$9/2$^+\#$&5.284&1&3.65$\times10^{8}$&2.35$\times10^{8}$&0.055\\
$^{223}$Fr$\to$$^{219}$At&3/2$^{(-)}$$\to$5/2$^-\#$&5.562&2&2.2$\times10^{7}$&1.71$\times10^{7}$&0.111\\
$^{223}$Ra$\to$$^{219}$Rn&3/2$^+$$\to$5/2$^+$&5.979&2&9.88$\times10^{5}$&4.12$\times10^{5}$&0.114\\
$^{223}$Ac$\to$$^{219}$Fr&(5/2$^-$)$\to$9/2$^-$&6.783&2&1.27$\times10^{2}$&2.16$\times10^{2}$&0.185\\
$^{223}$Th$\to$$^{219}$Ra&(5/2)$^+$$\to$(7/2)$^+$&7.566&2&6$\times10^{-1}$&1.08$\times10^{0}$&0.126\\
$^{223}$Pa$\to$$^{219}$Ac&9/2$^-\#$$\to$9/2$^-$&8.325&0&5.1$\times10^{-3}$&2.9$\times10^{-3}$&0.261\\
$^{225}$Ac$\to$$^{221}$Fr&3/2$^-\#$$\to$5/2$^-$&5.935&2&8.57$\times10^{5}$&1.33$\times10^{6}$&0.169\\
$^{225}$Th$\to$$^{221}$Ra&(3/2$^+$)$\to$5/2$^+$&6.922&2&5.83$\times10^{2}$&2.16$\times10^{2}$&0.134\\
$^{225}$U$\to$$^{221}$Th&5/2$^+\#$$\to$(7/2$^+$)&8.015&2&6.1$\times10^{-2}$&2.34$\times10^{-1}$&0.116\\
$^{225}$Np$\to$$^{221}$Pa&9/2$^-\#$$\to$9/2$^-$&8.785&0&3$\times10^{-3}$&5.85$\times10^{-4}$&0.33\\
$^{227}$Ac$\to$$^{223}$Fr&3/2$^-$$\to$3/2$^{(-)}$&5.042&0&4.98$\times10^{10}$&8.28$\times10^{10}$&0.144\\
$^{227}$Th$\to$$^{223}$Ra&1/2$^+$$\to$3/2$^+$&6.147&2&1.61$\times10^{6}$&5.3$\times10^{5}$&0.113\\
$^{227}$Pa$\to$$^{223}$Ac&(5/2$^-$)$\to$(5/2$^-$)&6.58&0&2.7$\times10^{3}$&4.77$\times10^{3}$&0.237\\
$^{227}$U$\to$$^{223}$Th&(3/2$^+$)$\to$(5/2)$^+$&7.211&2&6.6$\times10^{1}$&1.09$\times10^{2}$&0.145\\
$^{227}$Np$\to$$^{223}$Pa&5/2$^-\#$$\to$9/2$^-\#$&7.815&2&5.1$\times10^{-1}$&8.89$\times10^{-1}$&0.33\\
$^{229}$Th$\to$$^{225}$Ra&5/2$^+$$\to$1/2$^+$&5.168&2&2.5$\times10^{11}$&1.51$\times10^{11}$&0.075\\
$^{229}$Pa$\to$$^{225}$Ac&(5/2$^+$)$\to$3/2$^-\#$&5.834&1&2.7$\times10^{7}$&2.1$\times10^{7}$&0.206\\
$^{229}$U$\to$$^{225}$Th&(3/2$^+$)$\to$(3/2$^+$)&6.475&0&1.74$\times10^{4}$&6.22$\times10^{4}$&0.144\\
$^{229}$Np$\to$$^{225}$Pa&5/2$^+\#$$\to$5/2$^-\#$&7.015&1&3.53$\times10^{2}$&5.25$\times10^{2}$&0.309\\
$^{229}$Pu$\to$$^{225}$U&3/2$^+\#$$\to$5/2$^+\#$&7.587&2&1.82$\times10^{2}$&3.05$\times10^{1}$&0.145\\
$^{231}$Th$\to$$^{227}$Ra&5/2$^+$$\to$3/2$^+$&4.213&2&2.3$\times10^{17}$&2.27$\times10^{18}$&0.04\\
$^{231}$Pa$\to$$^{227}$Ac&3/2$^-$$\to$3/2$^-$&5.15&0&1.03$\times10^{12}$&1.63$\times10^{11}$&0.167\\
$^{231}$U$\to$$^{227}$Th&(5/2)$^{(+\#)}$$\to$1/2$^+$&5.577&2&9.07$\times10^{9}$&4.25$\times10^{9}$&0.112\\
$^{231}$Np$\to$$^{227}$Pa&(5/2)$^{(+\#)}$$\to$(5/2$^-$)&6.373&1&1.46$\times10^{5}$&3.21$\times10^{5}$&0.271\\
$^{231}$Pu$\to$$^{227}$U&3/2$^+\#$$\to$(3/2$^+$)&6.839&0&3.97$\times10^{3}$&1.11$\times10^{4}$&0.168\\
$^{233}$Pu$\to$$^{229}$U&5/2$^+\#$$\to$(3/2$^+$)&6.414&2&1.04$\times10^{6}$&1.46$\times10^{6}$&0.153\\
$^{233}$Am$\to$$^{229}$Np&5/2$^-\#$$\to$5/2$^+\#$&7.055&1&4.27$\times10^{3}$&2.25$\times10^{3}$&0.331\\
$^{233}$Cm$\to$$^{229}$Pu&3/2$^+\#$$\to$3/2$^+\#$&7.475&0&1.35$\times10^{2}$&2.19$\times10^{2}$&0.182\\
$^{235}$U$\to$$^{231}$Th&7/2$^-$$\to$5/2$^+$&4.678&1&2.22$\times10^{16}$&5.25$\times10^{15}$&0.034\\
$^{235}$Np$\to$$^{231}$Pa&5/2$^+$$\to$3/2$^-$&5.194&1&1.32$\times10^{12}$&1.09$\times10^{12}$&0.172\\
$^{235}$Pu$\to$$^{231}$U&(5/2$^+$)$\to$(5/2)$^{(+\#)}$&5.951&0&5.42$\times10^{7}$&2.05$\times10^{8}$&0.111\\
$^{235}$Am$\to$$^{231}$Np&5/2$^-\#$$\to$(5/2)$^{(+\#)}$&6.575&1&1.55$\times10^{5}$&2.81$\times10^{5}$&0.276\\
$^{237}$Np$\to$$^{233}$Pa&5/2$^+$$\to$3/2$^-$&4.959&1&6.76$\times10^{13}$&4.92$\times10^{13}$&0.125\\
$^{237}$Pu$\to$$^{233}$U&7/2$^-$$\to$5/2$^+$&5.748&1&9.39$\times10^{10}$&4.84$\times10^{9}$&0.063\\
$^{237}$Am$\to$$^{233}$Np&5/2$^{(-)}$$\to$5/2$^+\#$&6.195&1&1.77$\times10^{7}$&2.07$\times10^{7}$&0.216\\
$^{237}$Cm$\to$$^{233}$Pu&5/2$^+\#$$\to$5/2$^+\#$&6.775&0&6.67$\times10^{4}$&1.36$\times10^{5}$&0.164\\
$^{239}$Np$\to$$^{235}$Pa&5/2$^+$$\to$(3/2$^-$)&4.599&1&4.07$\times10^{16}$&2.7$\times10^{16}$&0.085\\
$^{239}$Am$\to$$^{235}$Np&(5/2)$^-$$\to$5/2$^+$&5.922&1&4.28$\times10^{8}$&6.54$\times10^{8}$&0.159\\
$^{239}$Cm$\to$$^{235}$Pu&(7/2$^-$)$\to$(5/2$^+$)&6.541&1&1.45$\times10^{8}$&2.51$\times10^{6}$&0.11\\
$^{241}$Pu$\to$$^{237}$U&5/2$^+$$\to$1/2$^+$&5.14&2&1.84$\times10^{13}$&1.65$\times10^{14}$&0.01\\
$^{241}$Am$\to$$^{237}$Np&5/2$^-$$\to$5/2$^+$&5.638&1&1.36$\times10^{10}$&3.22$\times10^{10}$&0.109\\
$^{243}$Am$\to$$^{239}$Np&5/2$^-$$\to$5/2$^+$&5.439&1&2.32$\times10^{11}$&6.8$\times10^{11}$&0.07\\
$^{243}$Cm$\to$$^{239}$Pu&5/2$^+$$\to$1/2$^+$&6.169&2&9.18$\times10^{8}$&8.49$\times10^{8}$&0.024\\
$^{243}$Bk$\to$$^{239}$Am&3/2$^-\#$$\to$(5/2)$^-$&6.874&2&1.08$\times10^{7}$&2.38$\times10^{5}$&0.13\\
$^{243}$Es$\to$$^{239}$Bk&(7/2$^+$)$\to$(7/2$^+$)&8.075&0&3.54$\times10^{1}$&1.58$\times10^{1}$&0.204\\
$^{243}$Fm$\to$$^{239}$Cf&7/2$^-\#$$\to$5/2$^+\#$&8.685&1&2.54$\times10^{-1}$&3.86$\times10^{-1}$&0.259\\
$^{245}$Cm$\to$$^{241}$Pu&7/2$^+$$\to$5/2$^+$&5.623&2&2.66$\times10^{11}$&2.19$\times10^{12}$&0.008\\
$^{245}$Bk$\to$$^{241}$Am&3/2$^-$$\to$5/2$^-$&6.454&2&3.56$\times10^{8}$&2.68$\times10^{7}$&0.085\\
$^{247}$Cm$\to$$^{243}$Pu&9/2$^-$$\to$7/2$^+$&5.353&1&4.92$\times10^{14}$&2.3$\times10^{14}$&0.002\\
$^{247}$Bk$\to$$^{243}$Am&(3/2$^-$)$\to$5/2$^-$&5.89&2&4.35$\times10^{10}$&3.08$\times10^{10}$&0.052\\
$^{247}$Cf$\to$$^{243}$Cm&7/2$^+\#$$\to$5/2$^+$&6.495&2&3.2$\times10^{7}$&1.97$\times10^{8}$&0.02\\
$^{247}$Fm$\to$$^{243}$Cf&(7/2$^+$)$\to$1/2$^+\#$&8.255&4&6.2$\times10^{1}$&8.72$\times10^{1}$&0.107\\
$^{247}$Fm$^m$$\to$$^{243}$Cf&(1/2$^+$)$\to$1/2$^+\#$&8.305&0&5.1$\times10^{0}$&1.07$\times10^{1}$&0.107\\
$^{247}$Md$\to$$^{243}$Es&(7/2$^-$)$\to$(7/2$^+$)&8.765&1&1.19$\times10^{0}$&7.78$\times10^{-1}$&0.147\\
$^{247}$Md$^m$$\to$$^{243}$Es&(1/2$^-$)$\to$(7/2$^+$)&9.025&3&3.16$\times10^{-1}$&3$\times10^{-1}$&0.147\\
$^{249}$Bk$\to$$^{245}$Am&7/2$^+$$\to$(5/2)$^+$&5.524&2&1.97$\times10^{12}$&6.35$\times10^{12}$&0.03\\
$^{249}$Cf$\to$$^{245}$Cm&9/2$^-$$\to$7/2$^+$&6.296&1&1.11$\times10^{10}$&3.85$\times10^{9}$&0.006\\
$^{251}$Es$\to$$^{247}$Bk&(3/2$^-$)$\to$(3/2$^-$)&6.598&0&2.38$\times10^{7}$&5.49$\times10^{7}$&0.035\\
$^{251}$Fm$\to$$^{247}$Cf&(9/2$^-$)$\to$7/2$^+\#$&7.425&1&1.06$\times10^{6}$&1.01$\times10^{5}$&0.017\\
$^{251}$Md$\to$$^{247}$Es&(7/2$^-$)$\to$(7/2$^+$)&7.964&1&2.53$\times10^{3}$&7.03$\times10^{2}$&0.062\\
$^{251}$No$\to$$^{247}$Fm&(7/2$^+$)$\to$(7/2$^+$)&8.755&0&9.64$\times10^{-1}$&1.99$\times10^{0}$&0.106\\
$^{253}$Md$\to$$^{249}$Es&(7/2$^-$)$\to$7/2$^+$&7.575&1&1.2$\times10^{5}$&3.05$\times10^{4}$&0.037\\
$^{253}$Lr$\to$$^{249}$Md&(7/2$^-$)$\to$(7/2$^-$)&8.925&0&7.02$\times10^{-1}$&2.33$\times10^{0}$&0.061\\
$^{255}$Lr$^m$$\to$$^{251}$Md&(7/2$^-$)$\to$(7/2$^-$)&8.594&0&2.54$\times10^{0}$&3.88$\times10^{1}$&0.036\\
$^{255}$Rf$^m$$\to$$^{251}$No&5/2$^+\#$$\to$(7/2$^+$)&8.975&2&1$\times10^{0}$&3.44$\times10^{0}$&0.105\\
$^{257}$Db$\to$$^{253}$Lr&(9/2$^+$)$\to$(7/2$^-$)&9.205&1&2.45$\times10^{0}$&3.46$\times10^{0}$&0.033\\
\end{longtable*}
\endgroup

\clearpage
\begingroup
\renewcommand*{\arraystretch}{1.3}
\begin{longtable*}{ccccccc}
\caption{Same as Table II, but for odd-$A$ nuclei in region IV.}\label{tab5} \\
\hline 
$\alpha$ transition & $I_i^\pi \to I_j^\pi$ & $Q_\alpha$(MeV) &$l_\text{min}$ & $T^\text{expt}_{1/2}$(s) & $T^\text{calc}_{1/2}$(s) & $P_\alpha$ \\ \hline
\endfirsthead
\multicolumn{7}{c}%
{{\tablename\ \thetable{} -- continued}} \\
\hline 
$\alpha$ transition & $I_i^\pi \to I_j^\pi$ & $Q_\alpha$(MeV) &$l_\text{min}$ & $T^\text{expt}_{1/2}$(s) & $T^\text{calc}_{1/2}$(s) & $P_\alpha$ \\ \hline
\endhead
\hline \\
\endfoot
\hline \hline
\endlastfoot
$^{253}$Es$\to$$^{249}$Bk&7/2$^+$$\to$7/2$^+$&6.739&0&1.77$\times10^{6}$&4.22$\times10^{6}$&0.087\\
$^{255}$Cf$\to$$^{251}$Cm&(7/2$^+$)$\to$(1/2$^+$)&5.736&4&2.55$\times10^{12}$&1.4$\times10^{12}$&0.064\\
$^{255}$Fm$\to$$^{251}$Cf&7/2$^+$$\to$1/2$^+$&7.24&4&7.23$\times10^{4}$&6.21$\times10^{5}$&0.059\\
$^{255}$Md$\to$$^{251}$Es&(7/2$^-$)$\to$(3/2$^-$)&7.905&2&2.03$\times10^{4}$&1.29$\times10^{3}$&0.061\\
$^{257}$Fm$\to$$^{253}$Cf&(9/2$^+$)$\to$(7/2$^+$)&6.864&2&8.68$\times10^{6}$&9.71$\times10^{6}$&0.042\\
$^{257}$Md$\to$$^{253}$Es&(7/2$^-$)$\to$7/2$^+$&7.558&1&1.32$\times10^{5}$&2.42$\times10^{4}$&0.042\\
$^{257}$Rf$^m$$\to$$^{253}$No&(11/2$^-$)$\to$(9/2$^-$)&9.156&2&4.3$\times10^{0}$&1.5$\times10^{0}$&0.059\\
$^{259}$Sg$^m$$\to$$^{255}$Rf$^m$&(9/2$^-$)$\to$5/2$^+\#$&9.885&2&3.11$\times10^{-1}$&6.25$\times10^{-2}$&0.061\\
$^{261}$Rf$\to$$^{257}$No&3/2$^+\#$$\to$(7/2$^+$)&8.645&2&8.15$\times10^{0}$&9.93$\times10^{1}$&0.029\\
$^{261}$Sg$\to$$^{257}$Rf&(3/2$^+$)$\to$(1/2$^+$)&9.713&2&1.87$\times10^{-1}$&2.45$\times10^{-1}$&0.042\\
$^{263}$Rf$\to$$^{259}$No&3/2$^+\#$$\to$9/2$^+\#$&8.255&4&2.2$\times10^{3}$&7.93$\times10^{3}$&0.022\\
$^{263}$Sg$\to$$^{259}$Rf&7/2$^+\#$$\to$7/2$^+\#$&9.405&0&1.08$\times10^{0}$&1.42$\times10^{0}$&0.03\\
$^{265}$Sg$^m$$\to$$^{261}$Rf&3/2$^+\#$$\to$3/2$^+\#$&9.125&0&2.46$\times10^{1}$&1.16$\times10^{1}$&0.023\\
$^{265}$Hs$\to$$^{261}$Sg&3/2$^+\#$$\to$(3/2$^+$)&10.47&0&1.96$\times10^{-3}$&7.5$\times10^{-3}$&0.037\\
$^{267}$Hs$\to$$^{263}$Sg&5/2$^+\#$$\to$7/2$^+\#$&10.035&2&6.88$\times10^{-2}$&1.93$\times10^{-1}$&0.028\\
$^{267}$Ds$\to$$^{263}$Hs&9/2$^+\#$$\to$7/2$^+\#$&11.775&2&1$\times10^{-5}$&3.51$\times10^{-5}$&0.053\\
$^{269}$Ds$\to$$^{265}$Hs&9/2$^+\#$$\to$3/2$^+\#$&11.514&4&2.3$\times10^{-4}$&4.46$\times10^{-4}$&0.041\\
$^{271}$Ds$\to$$^{267}$Hs&13/2$^-\#$$\to$5/2$^+\#$&10.875&5&9$\times10^{-2}$&3.21$\times10^{-2}$&0.033\\
$^{271}$Ds$^m$$\to$$^{267}$Hs&9/2$^+\#$$\to$5/2$^+\#$&10.945&2&1.7$\times10^{-3}$&3.41$\times10^{-3}$&0.033\\
$^{273}$Ds$\to$$^{269}$Hs&13/2$^-\#$$\to$9/2$^+\#$&11.365&3&2.4$\times10^{-4}$&6.22$\times10^{-4}$&0.028\\
$^{273}$Ds$^m$$\to$$^{269}$Hs&3/2$^+\#$$\to$9/2$^+\#$&11.565&4&1.2$\times10^{-1}$&4.12$\times10^{-4}$&0.028\\
$^{277}$Ds$\to$$^{273}$Hs&11/2$^+\#$$\to$3/2$^+\#$&10.835&4&2.2$\times10^{-2}$&1.9$\times10^{-2}$&0.025\\
$^{281}$Ds$\to$$^{277}$Hs&3/2$^+\#$$\to$3/2$^+\#$&9.325&0&9.33$\times10^{1}$&3.95$\times10^{1}$&0.027\\
$^{281}$Cn$\to$$^{277}$Ds&3/2$^+\#$$\to$11/2$^+\#$&10.465&4&3.7$\times10^{-1}$&3.73$\times10^{-1}$&0.043\\
$^{285}$Cn$\to$$^{281}$Ds&5/2$^+\#$$\to$3/2$^+\#$&9.315&2&3.2$\times10^{1}$&1.68$\times10^{2}$&0.053\\
$^{289}$Fl$\to$$^{285}$Cn&5/2$^+\#$$\to$5/2$^+\#$&9.965&0&2.4$\times10^{0}$&2.25$\times10^{0}$&0.131\\
\end{longtable*}
\endgroup

\end{document}